\documentclass[12pt]{article}
\usepackage{a4,latexsym, amssymb}

\newtheorem{Theorem}{Theorem}

\newtheorem{Corollary}{Corollary}
\newtheorem{Proposition}{Proposition}
\newtheorem{Definition}{Definition}

\newtheorem{Remark}{Remark}

\begin{document}

\title{Johnson Type Bounds on Constant Dimension Codes
\thanks{This research is supported in part by the
NSFC-GDSF Joint Fund under Grant No. U0675001, and the open
research fund of National Mobile Communications Research
Laboratory, Southeast University. }}

\author{
Shu-Tao Xia\thanks{S.-T. Xia is with the Graduate School at Shenzhen
of Tsinghua University, Shenzhen, Guangdong 518055, P. R. China. He
is also with the National Mobile Communications Research Laboratory,
Southeast University, P.R. China. E-mail: xiast@sz.tsinghua.edu.cn}
\ and Fang-Wei Fu\thanks{F.-W. Fu is with the Chern Institute of
Mathematics, and The Key Laboratory of Pure Mathematics and
Combinatorics, Nankai University, Tianjin 300071, P.R. China. Email:
fwfu@nankai.edu.cn}}

\date{}
\maketitle

\begin{abstract}
Very recently, an operator channel was defined by Koetter and
Kschischang when they studied random network coding. They also
introduced constant dimension codes and demonstrated that these
codes can be employed to correct errors and/or erasures over the
operator channel. Constant dimension codes are equivalent to the
so-called linear authentication codes introduced by Wang, Xing and
Safavi-Naini when constructing distributed authentication systems
in 2003. In this paper, we study constant dimension codes. It is
shown that Steiner structures are optimal constant dimension codes
achieving the Wang-Xing-Safavi-Naini bound. Furthermore, we show
that constant dimension codes achieve the Wang-Xing-Safavi-Naini
bound if and only if they are certain Steiner structures. Then, we
derive two Johnson type upper bounds, say I and II, on constant
dimension codes. The Johnson type bound II slightly improves on
the Wang-Xing-Safavi-Naini bound. Finally, we point out that a
family of known Steiner structures is actually a family of optimal
constant dimension codes achieving both the Johnson type bounds I
and II.
\end{abstract}

{\bf keywords:}\quad Constant dimension codes, linear
authentication codes, binary constant weight codes, Johnson
bounds, Steiner structures, random network coding.

\baselineskip=18pt

\section{Introduction}

Throughout this paper, $\mathbb{F}_q$ denotes the finite field
with $q$ elements, where $q$ is a prime power. Let $W$ be an
$n$-dimensional vector space over $\mathbb{F}_q$ and let
$\mathcal{P}(W)$ denote the set of all subspaces of $W$. For any
$A, B \in \mathcal{P}(W)$, denote $$A+B=\{a+b:a\in A, b\in B\},$$
that is the smallest subspace containing both $A$ and $B$. It is
known \cite{kk} that the \emph{dimension distance} between $A$ and
$B$ defined by
\begin{eqnarray}
\label{d1}
d(A,B)&=& \dim(A+B)- \dim(A \cap B)  \\
\label{d2} &=& \dim(A) + \dim(B) - 2 \dim(A \cap B)
\end{eqnarray}
is a metric for the space $\mathcal{P}(W)$. A $q$-ary $(n,M,D)$ or
$(n,M,D)_q$ code $\mathcal{C}$ is simply a subset of
$\mathcal{P}(W)$ with size $M$ and \emph{minimum dimension
distance} $D$ which is defined by
\begin{eqnarray}
D=D(\mathcal{C})&=&\min_{X\ne Y\in \mathcal{C}} d(X,Y).
\end{eqnarray}
For any positive integer $l\le n$, let $\mathcal{P}(W,l)$ denote
the set of all $l$-dimensional subspaces of $W$. For integers
$0\le m\le n$ and $q\ge 2$, let
$$\left[n\atop m\right]_q = \prod_{i=0}^{m-1}
\frac{q^{n-i}-1}{q^{m-i}-1}$$ denote the $q$-binomial coefficient
or \emph{Gaussian binomial coefficient} \cite[pp.443-444]{ms}. It
is well known that $|\mathcal{P}(W,l)|= \left[n\atop l\right]_q$.
A $q$-ary $(n,M,2\delta,l)$ or $(n,M,2\delta,l)_q$ constant
dimension code is simply a subset of $\mathcal{P}(W,l)$ with size
$M$ and minimum dimension distance $2\delta$. Note that by
(\ref{d2}) the dimension distance of any two codewords of a
constant dimension code must be an even number and $1 \le \delta
\le l$. An $(n,M,\geq 2\delta,l)_q$ constant dimension code is a
subset of $\mathcal{P}(W,l)$ with size $M$ and minimum dimension
distance at least $2\delta$. For fixed numbers $n,l,\delta, q$,
denote $A_q[n,2\delta,l]$ the maximum number $M$ of codewords in
an $(n,M,\geq 2\delta,l)_q$ constant dimension code. An $(n,M,\geq
2\delta,l)_q$ constant dimension code is said to be \emph{optimal}
if $M=A_q[n,2\delta,l]$. One of the main research problems on
constant dimension codes is to determine $A_q[n,2\delta,l]$ and
find corresponding optimal constant dimension codes.

Denote $X^\perp$ the orthogonal complement of $X\in
\mathcal{P}(W)$. For any two $l$-dimensional subspaces $X,Y\in
\mathcal{P}(W,l)$, since $X^\perp\cap Y^\perp=(X+Y)^\perp$, we
have
\begin{eqnarray}
d(X^\perp,Y^\perp)&=& {\rm dim}(X^\perp)+{\rm dim}(Y^\perp)-2{\rm
dim}(X^\perp\cap Y^\perp)\nonumber\\
&=& n-{\rm dim}(X)+n-{\rm dim}(Y)-2(n-{\rm
dim}(X+Y))\nonumber\\
&=& d(X,Y).\label{m0}
\end{eqnarray}
Let $\mathcal{C}\subseteq \mathcal{P}(W,l)$ be an
$(n,M,2\delta,l)_q$ constant dimension code. Then by (\ref{m0}) we
know that $\bar\mathcal{C}\triangleq \{X^\perp: X\in
\mathcal{C}\}$ is an $(n,M,2\delta,n-l)_q$ constant dimension
code. This implies that
\begin{eqnarray}
A_q[n,2\delta,l]=A_q[n,2\delta,n-l]. \label{fw1}
\end{eqnarray}
Hence, we only need to determine $A_q[n,2\delta,l]$ for $l\leq
n/2$.

When studying random network coding \cite{hkmke,hmkkesl}, Koetter
and Kschischang \cite{kk} defined a so-called \emph{operator
channel} and found that an $(n,M,\geq 2\delta,l)_q$ constant
dimension code $\mathcal{C}$ could be employed to correct errors
and/or erasures over the operator channel, i.e., the errors and/or
erasures could be corrected by a minimum dimension distance
decoder if the sum of errors and erasures is less than $\delta$.
Some bounds on $A_q[n,2\delta,l]$, e.g., the Hamming type upper
bound, the Gilbert type lower bound, and the Singleton type upper
bound, were derived in \cite{kk}. It is known that the Hamming
type bound is not very good \cite{kk} and there exist no
non-trivial perfect codes meeting the Hamming type bound
\cite{se,e}. The Singleton type bound developed in \cite{kk} is
the following:
\begin{Proposition}
\label{prop-skk} {\rm \cite[Th.3]{kk}\quad (Singleton type bound)}
$$A_q[n,2\delta,l]\le \left[n-\delta+1\atop l-\delta+1\right]_q.$$
\end{Proposition}
Moreover, Koetter and Kschischang  \cite{kk} designed a class of
Reed-Solomon like constant dimension codes and afforded decoding
procedures. They showed that these codes were nearly
Singleton-type-bound-achieving.

In 2003, Wang, Xing and Safavi-Naini \cite{wxs} introduced the
so-called \emph{linear authentication codes} when constructing
distributed authentication systems. They \cite[Th.4.1]{wxs} showed
that an $(n,M,\geq 2\delta,l)_q$ constant dimension code is
exactly an $[n,M,t=n-l,d=\delta]$ linear authentication code over
$\mathbb{F}_q$. Furthermore, they established an upper bound
\cite[Th.5.2]{wxs} on linear authentication codes, which is
equivalent to the following bound on constant dimension codes:

\begin{Proposition}
\label{prop-wxs} {\rm \cite[Th.5.2]{wxs}\quad
(Wang-Xing-Safavi-Naini Bound)}
$$A_q[n,2\delta,l]\le \frac{\left[n\atop
l-\delta+1\right]_q}{\left[l\atop l-\delta+1\right]_q}.$$
\end{Proposition}
Moreover, Wang, Xing and Safavi-Naini \cite{wxs} presented some
constructions of linear authentication codes (or corresponding
constant dimension codes) that are asymptotically close to this
bound.

In this paper, we show that Steiner structures are optimal
constant dimension codes achieving the Wang-Xing-Safavi-Naini
bound in Proposition \ref{prop-wxs}. Furthermore, it is shown that
constant dimension codes achieve the Wang-Xing-Safavi-Naini bound
if and only if they are certain Steiner structures. Two Johnson
type upper bounds, say I and II, on constant dimension codes are
derived. The Johnson type bound II slightly improves on the
Wang-Xing-Safavi-Naini bound. It is observed that the
Wang-Xing-Safavi-Naini bound is always better than the Singleton
type bound for nontrivial constant dimension codes. Finally, we
point out that a family of known Steiner structures is actually a
family of optimal constant dimension codes achieving both the
Johnson type bounds I and II.

\section{Steiner Structures}

In this section we first introduce the combinatorial objectives
Steiner structures. Then we show that constant dimension codes
achieve the Wang-Xing-Safavi-Naini bound if and only if they are
certain Steiner structures. This means that Steiner structures are
optimal constant dimension codes. Finally we describe the only
known family of nontrivial Steiner structures in combinatorics.

Recall that $W$ is the $n$-dimensional vector space over the
finite field $\mathbb{F}_q$ and $\mathcal{P}(W,l)$ denote the set
of all $l$-dimensional subspaces of $W$. The following definition
and proposition on Steiner structures are from \cite{se}.

\begin{Definition}
\label{def1} {\rm\cite{se}} A subset $\mathcal{F}\subseteq
\mathcal{P}(W,l)$ is called a \emph{Steiner structure}
$S[t,l,n]_q$ if each $t$-dimensional subspace of $W$ is contained
in exactly one $l$-dimensional subspace from $\mathcal{F}$. The
$l$-dimensional subspaces in $\mathcal{F}$ are called blocks of
the Steiner structure $S[t,l,n]_q$.
\end{Definition}

\begin{Proposition}{\rm\cite{se}}
\label{prop-s1} The total number of blocks in an $S[t,l,n]_q$ is
$\left[n\atop t\right]_q /\left[l\atop t\right]_q$.
\end{Proposition}

Below we show that Steiner structures are constant dimension
codes.
\begin{Proposition}
\label{prop-Steiner} A Steiner structure $S[t,l,n]_q$ is an
$(n,M,2\delta,l)_q$ constant dimension code with $M=\left[n\atop
t\right]_q /\left[l\atop t\right]_q$ and $\delta=l-t+1$.
\end{Proposition}
\begin{proof}
By Definition \ref{def1} and Proposition \ref{prop-s1}, we only
need to show that $\delta=l-t+1$. For any two different blocks
$X,Y\in S[t,l,n]_q$, since every $t$-dimensional subspace is
contained in exactly one block of $S[t,l,n]_q$, we have ${\rm dim}
(X\cap Y) \le t-1$. Thus, by (\ref{d2}), $d(X,Y)=2l-2\;{\rm dim}
(X\cap Y)\ge 2(l-t+1)$, which implies that $\delta \ge l-t+1$. On
the other hand, let $V$ be a fixed $(t-1)$-dimensional subspace of
$W$, choose two $t$-dimensional subspaces $U_1$ and $U_2$ of $W$
such that $V=U_1\cap U_2$. Let $X_1$ and $X_2$ be the unique
blocks in $S[t,l,n]_q$ such that $U_1\subseteq X_1$ and
$U_2\subseteq X_2$, respectively. Then, $V\subseteq X_1\cap X_2$,
which implies that ${\rm dim} (X_1\cap X_2) \geq {\rm dim}
(V)=t-1$. Hence, by (\ref{d2}), $d(X_1,X_2)\le 2(l-t+1)$. Thus,
$\delta \le l-t+1$ since $2\delta$ is the minimum dimension
distance of $S[t,l,n]_q$. Combining these assertions,
$\delta=l-t+1$. This completes the proof.
\end{proof}

Next we give the necessary and sufficient condition for constant
dimension codes to achieve the Wang-Xing-Safavi-Naini bound in
Proposition \ref{prop-wxs}.

\begin{Theorem}\label{th-f1}
An $(n,M,\geq 2\delta,l)_q$ constant dimension code $\mathcal{C}$
achieves the Wang-Xing-Safavi-Naini bound, i.e.,
$M=\frac{\left[n\atop l-\delta+1\right]_q}{\left[l\atop
l-\delta+1\right]_q}$, if and only if $\mathcal{C}$ is a Steiner
structure $S[l-\delta+1,l,n]_q$.
\end{Theorem}
\begin{proof}
Since an $(n,M,2\delta,l)_q$ constant dimension code is an
$(n,M,\geq 2\delta,l)_q$ constant dimension code, we know from
Propositions \ref{prop-wxs} and \ref{prop-Steiner} that a Steiner
structure $S[l-\delta+1,l,n]_q$ is an $(n,M=\frac{\left[n\atop
l-\delta+1\right]_q}{\left[l\atop l-\delta+1\right]_q},\geq
2\delta,l)_q$ constant dimension code achieving the
Wang-Xing-Safavi-Naini bound.

On the other hand, suppose there exists an $(n,M=\frac{\left[n\atop
l-\delta+1\right]_q}{\left[l\atop l-\delta+1\right]_q},\geq
2\delta,l)_q$ constant dimension code $\mathcal{C}$ achieving the
Wang-Xing-Safavi-Naini bound. Since the dimension distance between
any two different codewords of $\mathcal{C}$ is not small than
$2\delta$, it follows from (\ref{d2}) that each
$(l-\delta+1)$-dimensional subspace could not be contained in two
different codewords. Moreover, since each codeword of $\mathcal{C}$
contains ${\left[l\atop l-\delta+1\right]_q}$ distinct
$(l-\delta+1)$-dimensional subspaces, all codewords of $\mathcal{C}$
contains totally $M{\left[l\atop l-\delta+1\right]_q}={\left[n\atop
l-\delta+1\right]_q}$ pairwisely different
$(l-\delta+1)$-dimensional subspaces. Note that there are totally
${\left[n\atop l-\delta+1\right]_q}$ distinct
$(l-\delta+1)$-dimensional subspaces of $W$. Hence, each
$(l-\delta+1)$-dimensional subspace is contained in exactly one
codeword of $\mathcal{C}$. Therefore, regarding the codewords of
$\mathcal{C}$ as blocks, $\mathcal{C}$ forms a Steiner structure
$S[l-\delta+1,l,n]_q$  by its definition.
\end{proof}

Theorem \ref{th-f1} shows that Steiner structures are optimal
constant dimension codes. The following corollary follows from
Theorem \ref{th-f1} imediately.
\begin{Corollary}\label{co-fw1}
$$A_q[n,2\delta,l]=\frac{\left[n\atop
l-\delta+1\right]_q}{\left[l\atop l-\delta+1\right]_q}
$$ if and only if a Steiner
structure $S[l-\delta+1,l,n]_q$ exists.
\end{Corollary}

It is known $\cite{se,e}$ that trivial Steiner structures
$S[t,n,n]_q$ and $S[t,t,n]_q$ exist for all $t\le n$. For
nontrivial Steiner structures, by our knowledge, it is only known
$\cite{se,e}$ that $S[1,l,n]_q$ exists where $l\mid n$, and the
blocks of $S[1,l,n]_q$ form a partition of $W$ (excluding the zero
vector). For completeness, we review the construction
$\cite{se,e}$ of such an $S[1,l,n]_q$ where $n=kl$ as follows. Let
$e=(q^{kl}-1)/(q^l-1)$ and let $\alpha$ be a primitive element of
$\mathbb{F}_{q^n}$. Define
$$\langle \alpha^e\rangle =\{1, \alpha^e,
\alpha^{2e},\ldots,\alpha^{(q^l-2)e}\}.$$ The cosets
$$C_i=\alpha^i\langle \alpha^e\rangle =\{\alpha^{i}, \alpha^{i+e},
\alpha^{i+2e},\ldots,\alpha^{i+(q^l-2)e}\}, \ i=0,1,\ldots,e-1 $$
are called {\em cyclotomic classes} of order $e$. Let
$E_i=C_i\cup\{0\}$. Note that $\mathbb{F}_{q^n}$ is an
$n$-dimensional vector space over $\mathbb{F}_{q}$. One can verify
that $E_0=\mathbb{F}_{q^l}$ and the $E_i$'s, when viewed as
subsets of $\mathbb{F}_q^n$, are $l$-dimensional subspaces of
$W=\mathbb{F}_q^n$. Regarding all $E_i$'s as blocks, an
$S[1,l,kl]_q$ is obtained.

From Corollary \ref{co-fw1}, we have the following result.
\begin{Corollary}\label{co-fw2}
\label{ths} For any positive integers $k$ and $l$, we
have
\begin{eqnarray}
A_q[kl,2l,l]=\frac{q^{kl}-1}{q^l-1}.\label{eq-fw2}
\end{eqnarray}
\end{Corollary}

\section{Johnson Type Bound I}

In this section, we first review some basic definitions and the
Johnson bound I for binary constant weight codes in coding theory.
It is shown that a corresponding binary constant weight code can
be obtained from a given constant dimension code. Then, using the
Johnson bound I for this corresponding binary constant weight
code, we obtain the Johnson type bound I for constant dimension
codes. It is observed that this bound is tight in some cases.

Let $\mathbb{F}_2^n$ be the $n$-dimensional vector space over the
binary field $\mathbb{F}_{2}$. For any two vectors $\mathbf{a},
\mathbf{b}\in \mathbb{F}_2^n$, the Hamming distance
$d_{H}(\mathbf{a}, \mathbf{b})$ is the number of coordinates in
which they differ, the Hamming weight $w_{H}(\mathbf{a})$ is the
number of nonzero coordinates in $\mathbf{a}$. It is known that
\begin{eqnarray}
d_{H}(\mathbf{a}, \mathbf{b})=
w_{H}(\mathbf{a})+w_{H}(\mathbf{b})-2w_{H}(\mathbf{a}*\mathbf{b})\label{eq-fw3}
\end{eqnarray}
where $$\mathbf{a}*\mathbf{b}=(a_1b_1,a_2b_2,\ldots,a_nb_n).$$ A
binary code $C$ of length $n$ is a nonempty subset of
$\mathbb{F}_2^n$. The minimum distance of $C$ is the minimum
Hamming distance between any two distinct codewords in $C$. A
binary constant weight code is a binary code such that every
codeword has a fixed Hamming weight. Denote $A(n,2\delta,w)$ the
maximum number of codewords in a binary constant weight code with
length $n$, weight $w$ and minimum distance at least $2\delta$. We
state the Johnson bound I for binary constant weight codes in the
following proposition.

\begin{Proposition}
\label{prop-j1} {\rm(Johnson bound I) \cite{t}} \quad If
$w^2>n(w-\delta)$, then
$$A(n,2\delta,w)\le \left\lfloor\frac{n\delta}{w^2-n(w-\delta)}\right\rfloor,$$
where $\lfloor\cdot\rfloor$ denotes the floor function.
\end{Proposition}

Below we show that a corresponding binary constant weight code can
be obtained from a given constant dimension code.

Recall that $W$ is the $n$-dimensional vector space over the finite
field $\mathbb{F}_q$ and $\mathcal{P}(W,l)$ denote the set of all
$l$-dimensional subspaces of $W$. Let $\mathbf{0}$ denote the
all-zero vector in $W$ and $W^*=W\setminus \{\mathbf{0}\}$. Denote
$N=q^n-1$. Suppose all the vectors in $W^*$ are ordered from $1$ to
$N$. Define the incidence vector of a subset $X\subseteq W$ by
$$\mathbf{v}_X=(v_1,v_2,\ldots,v_N)\in \mathbb{F}_2^N$$
where $v_i=1$ if the $i$-th vector of $W^*$ is contained in $X$,
and $v_i=0$ otherwise. For any two $l$-dimensional subspaces
$X,Y\in \mathcal{P}(W,l)$, by (\ref{eq-fw3}) it is easy to see
that
\begin{eqnarray}
w_H(\mathbf{v}_X)&=& w_H(\mathbf{v}_Y)=q^l-1,\label{m1}\\
w_H(\mathbf{v}_X*\mathbf{v}_Y)&=& q^{{\rm dim}(X\cap Y)}-1,\label{m2}\\
d_H(\mathbf{v}_X,\mathbf{v}_Y)&=& 2(q^l-q^{{\rm dim}(X\cap
Y)}).\label{m3}
\end{eqnarray}
Let $\mathcal{C}$ be an $(n,M,2\delta,l)_q$ constant dimension code.
By (\ref{m1}), the incidence vectors of the codewords in
$\mathcal{C}$ form a binary constant weight code $\mathbf{C}$, which
is called the \emph{derived binary constant weight code of
$\mathcal{C}$}. From (\ref{m1}), (\ref{m3}) and the definition of
constant dimension codes, we have the following result.
\begin{Proposition}
\label{prop-cwc} Let $\mathcal{C}$ be an $(n,M,2\delta,l)_q$
constant dimension code. Then its derived binary constant weight
code $\mathbf{C}$ has the following parameters: length $N=q^n-1$,
size $M$, minimum distance $2(q^{l}-q^{l-\delta})$, and weight
$q^l-1$.
\end{Proposition}

Although every constant dimension code corresponds to a binary
constant weight code, the reverse proposition may not hold. Given a
binary $(q^n-1,M,2(q^l-q^{l-\delta}), q^l-1)$ constant weight code,
since its codewords may not be the incidence vectors of any
subspaces, the code may not correspond to any $(n,M,2\delta,l)_q$
constant dimension code. From Propositions \ref{prop-j1} and
\ref{prop-cwc} we obtain the Johnson type bound I for constant
dimension codes.
\begin{Theorem}{\rm (Johnson type
bound I for constant dimension
codes)}\label{th-j1} \\
If $(q^l-1)^2>(q^n-1)(q^{l-\delta}-1)$, then
$$A_q[n,2\delta,l]\le
\left\lfloor\frac{(q^l-q^{l-\delta})(q^n-1)}{(q^l-1)^2-(q^n-1)(q^{l-\delta}-1)}\right\rfloor.$$
\end{Theorem}

The Johnson type bound I for constant dimension codes is tight in
some cases. By Proposition \ref{prop-Steiner} and Corollary
\ref{co-fw2}, the Steiner structure $S[1,l,kl]_q$ is a
$(kl,\frac{q^{kl}-1}{q^l-1},2l,l)_q$ constant dimension code
achieving the Johnson type bound I for constant dimension codes.

\begin{Remark}
{\em There is another method to obtain a binary constant weight
code from a constant dimension code. Let $\tilde W$ be the set of
all $1$-dimensional subspaces of $W$. Denote $\tilde
N=\frac{q^n-1}{q-1}$. We can regard $\tilde W$ as $PG(n-1,q)$, the
$(n-1)$-dimensional projective geometry over $\mathbb{F}_{q}$ with
$\tilde N$ points \cite[Appendix B]{ms}, where each $X\in
\mathcal{P}(W,l)$ corresponds to an $(l-1)$-flat of $PG(n-1,q)$,
say $\tilde X$. Suppose all points in $PG(n-1,q)$ are ordered from
$1$ to $\tilde N$. Define the \emph{punctured incidence vector} of
$X\in \mathcal{P}(W,l)$ as the incidence vector of $\tilde X\in
PG(n-1,q)$. By putting together the punctured incidence vectors of
all codewords of an $(n,M,2\delta,l)_q$ constant dimension code
$\mathcal{C}$, we obtain a corresponding binary constant weight
code $\tilde\mathbf{C}$ which has length $\tilde
N=\frac{q^n-1}{q-1}$, size $M$, minimum distance
$\frac{2(q^{l}-q^{l-\delta})}{q-1}$, and weight
$\frac{q^l-1}{q-1}$. Note that the derived binary constant weight
code $\mathbf{C}$ can be obtained by concatenating $(q-1)$ times
of $\tilde\mathbf{C}$. Hence, we have
\begin{eqnarray}
\label{Aq} A_q[n,2\delta,l]\le
A\left(\frac{q^n-1}{q-1},\frac{2(q^{l}-q^{l-\delta})}{q-1},\frac{q^l-1}{q-1}\right).
\end{eqnarray}
However, by employing the Johnson bound I for binary constant weight
codes, (\ref{Aq}) implies the same results with Theorem
\ref{th-j1}.}
\end{Remark}

\section{Johnson Type Bound II}

In this section, we derive an upper bound for constant dimension
codes. We call this upper bound the Johnson type bound II for
constant dimension codes since it is similar to the Johnson bound
II for binary constant weight codes \cite{t}. The Johnson type
bound II for constant dimension codes slightly improves on the
Wang-Xing-Safavi-Naini bound.

Let $V_1,V_2\in \mathcal{P}(W)$ and $V_2\subseteq V_1$. Define
$$(V_2|V_1)^\perp=\{\mathbf{a}\in V_1: \forall\; \mathbf{b}\in V_2,\;\;
\mathbf{a}\mathbf{b}^T=0\},$$ i.e., $(V_2|V_1)^\perp$ is the
orthogonal complement of $V_2$ in $V_1$. For any $S\subseteq W$,
denote $\langle S\rangle$ the minimum subspace containing $S$.

\begin{Theorem}\label{th2}
$$A_q[n,2\delta,l]\le
\left\lfloor\frac{q^n-1}{q^l-1}A_q[n-1,2\delta,l-1]\right\rfloor.$$
\end{Theorem}
\begin{proof}
Suppose $\mathcal{C}$ is an optimal $(n,M,\geq2\delta,l)_q$ constant
dimension code with $M=A_q[n,2\delta,l]$. Consider the binary
$M\times (q^n-1)$ matrix, say $\mathcal{P}$, whose rows consist of
all the codewords of $\mathbf{C}$, where $\mathbf{C}$ is the derived
binary constant weight code of $\mathcal{C}$ in Proposition
\ref{prop-cwc}. Denote $\Delta$ the total number of 1's in the
matrix $\mathcal{P}$. Since each codeword of $\mathbf{C}$ has weight
$q^l-1$,
\begin{eqnarray}
\Delta=M(q^l-1)=A_q[n,2\delta,l](q^l-1).\label{Delta}
\end{eqnarray}

On the other hand, we will show that the number of 1's in each
column of $\mathcal{P}$ is not greater than $A_q[n-1,2\delta,l-1]$.
Recall that the positions of the incidence vectors are indexed by
the non-zero vectors of $W$. Without loss of generality, suppose
$\alpha_1\in W$ is the non-zero vector which indexes the first
column of $\mathcal{P}$. Let $$\mathcal{C}_1=\{X\in \mathcal{C} :
\mbox{the first component of $\mathbf{v}_X$ is 1}\}.$$ Hence, the
weight of the column indexed by $\alpha_1$ equals $|\mathcal{C}_1|$.
Noting that $\alpha_1\in X$ for any $X\in \mathcal{C}_1$, let
$\mathcal{C}_1'=\{(\langle \alpha_1\rangle |X)^\perp :X\in
\mathcal{C}_1\}$ and $W_1=(\langle \alpha_1\rangle |W)^\perp$.
Clearly, $W_1$ is an $(n-1)$-dimensional vector space over
$\mathbb{F}_{q}$ and each element of $\mathcal{C}_1'$ is an
$(l-1)$-dimensional subspace of $W_1$. Hence,
$\mathcal{C}_1'\subseteq \mathcal{P}(W_1,l-1)$ is a $q$-ary constant
dimension code with length $n-1$, size $|\mathcal{C}_1|$, and
dimension $l-1$. Moreover, for any two different codewords of
$\mathcal{C}_1'$, e.g., $(\langle \alpha_1\rangle |X)^\perp$ and
$(\langle \alpha_1\rangle |Y)^\perp$, where $X\ne Y\in
\mathcal{C}_1$,
\begin{eqnarray*}
&& d((\langle \alpha_1\rangle |X)^\perp ,(\langle \alpha_1\rangle
|Y)^\perp )\\
&=& 2(l-1)-2{\rm dim}[(\langle \alpha_1\rangle |X)^\perp \cap (\langle \alpha_1\rangle |Y)^\perp ]\\
&=& 2(l-1)-2{\rm dim}((\langle \alpha_1\rangle |X\cap Y)^\perp )\\
&=& 2l-2{\rm dim}(X\cap Y)\\
&=& d(X,Y) \ge 2\delta.
\end{eqnarray*}
Hence, $\mathcal{C}_1'$ is an
$(n-1,|\mathcal{C}_1|,\geq2\delta,l-1)_q$ constant dimension code,
which implies that $$|\mathcal{C}_1|\le A_q[n-1,2\delta,l-1].$$ The
weight of the column indexed by $\alpha_1$ is not greater than
$A_q[n-1,2\delta,l-1]$. Therefore, by counting the number of 1's for
each column of $\mathcal{P}$, we have that $$\Delta\le (q^n-1)
A_q[n-1,2\delta,l-1].$$ Combining this with (\ref{Delta}) and noting
that $A_q[n,2\delta,l]$ is an integer, we obtain the required
conclusion.
\end{proof}

Using Theorem \ref{th2} recursively, we obtain the Johnson type
bound II for constant dimension codes.

\begin{Corollary}\label{co-j2} {\rm (Johnson type bound II for constant dimension
codes)}
$$
A_q[n,2\delta,l]\le
\left\lfloor\frac{q^n-1}{q^l-1}\left\lfloor\frac{q^{n-1}-1}{q^{l-1}-1}
\left\lfloor\cdots\left\lfloor\frac{q^{n-l+\delta}-1}{q^\delta-1}\right\rfloor
\cdots\right\rfloor\right\rfloor\right\rfloor.$$
\end{Corollary}

The Johnson type bound II slightly improves on the
Wang-Xing-Safavi-Naini bound. Let $B_S, B_{WXS}, B_J$ denote
respectively the Singleton type bound in Proposition
$\ref{prop-skk}$, the Wang-Xing-Safavi-Naini bound in Proposition
$\ref{prop-wxs}$, and the Johnson type bound II in Corollary
$\ref{co-j2}$. For example, letting $q=2$, $n=6$, $\delta=2$ and
$l=3$, we have $B_S=155$, $B_{WXS}=93$ and $B_J=90$. Below we show
that the Wang-Xing-Safavi-Naini bound is always better than the
Singleton type bound for $\delta>1$ and $n>l$. Since for
$i=0,1,\ldots, l-\delta$,
\begin{eqnarray*}
\frac{(q^{n-l+i+1}-1)}{(q^{i+1}-1)} \ge
\frac{(q^{n-l+\delta+i}-1)}{(q^{\delta+i}-1)}&\Longleftrightarrow&
(q^{n-l+i+1}-q^{i+1})(q^{\delta-1}-1)\ge 0,
\end{eqnarray*}
we have that
\begin{eqnarray}
B_S &=& \frac{(q^{n-\delta+1}-1)(q^{n-\delta}-1)\cdots
(q^{n-l+1}-1)}{(q^{l-\delta+1}-1)(q^{l-\delta}-1)\cdots (q-1)}\nonumber\\
&\ge& \frac{(q^n-1)(q^{n-1}-1)\cdots
(q^{n-l+\delta}-1)}{(q^l-1)(q^{l-1}-1)\cdots
(q^\delta-1)}=B_{WXS}\label{com1}
\end{eqnarray}
and the equality holds if and only if $\delta=1$ or $n=l$.
Furthermore, by \cite[Lemma 5]{kk}, $1<q^{-l(n-l)}\left[n\atop
l\right]_q<4$ for $0<l<n$. Using the similar arguments in the proof
of \cite[Lemma 5]{kk}, we obtain
\begin{eqnarray}
\label{com0} 1<q^{-m(u-v)}\frac{(q^{u}-1)(q^{u-1}-1)\cdots
(q^{u-m+1}-1)}{(q^{v}-1)(q^{v-1}-1)\cdots (q^{v-m+1}-1)} <4 \;\mbox{
for $1\le m\le v< u$}.
\end{eqnarray}
Hence, by (\ref{com1}) and (\ref{com0}), it is easy to see that for
$\delta>1$ and $0<l<n$
\begin{eqnarray}
\label{com2} B_{WXS}<B_S<4 q^{(l-\delta+1)(n-l)}<4{B_{WXS}}.
\end{eqnarray}
For example, letting $n=100$, $l/n=0.4$, $\delta/n=0.2$, it is
computed with the \emph{Mathematica} software that
\begin{eqnarray*}
\frac{B_S}{B_{WXS}}&\approx& 3.46, 1.79, 1.45, 1.32, \quad\mbox{for
} q=2,3,4,5, \mbox{ respectively.}
\end{eqnarray*}

\section{Concluding Remarks}

In this paper we show that Steiner structures, e.g.,
$S[1,l,kl]_q$, are optimal constant dimension codes or linear
authentication codes, and could be applied in random network
coding or distributed authentication systems. Furthermore, it is
shown that constant dimension codes achieve the
Wang-Xing-Safavi-Naini bound if and only if they are certain
Steiner structures. We derive two Johnson type bounds for constant
dimension codes. It would be interesting to construct more
constant dimension codes which achieve Johnson type bounds I or
II. It is a hard problem to determine $A_q[n,2\delta,l]$ in
general. However, one can first make efforts to determine
$A_q[n,4,l]$, $A_q[n,6,l]$ and $A_q[n,2(l-1),l]$ in the following
steps.

\section*{Acknowledgment}
The authors would like to thank Professor Cunsheng Ding for
reading this paper and giving valuable comments that helped to
improve the paper.

\end{document}